\documentclass{article}
\usepackage[margin=0.8in]{geometry}
\usepackage{amsmath}
\usepackage{authblk}
\usepackage{gensymb}
\usepackage{xr}
\usepackage{graphicx}

\author[1]{Marcus Tamura}
\author[3,4]{Chuanyu Lian}
\author[3,4]{Hongyi Sun}
\author[3,4]{Yi-Siou Huang}
\author[2]{Sadra Rahimi Kari}
\author[1]{Zhimu Guo}
\author[1]{Alexander N. Tait}
\author[1]{Nir Rotenberg}
\author[2]{Nathan Youngblood}
\author[3,4]{Carlos A. R\'{i}os Ocampo}
\author[1]{Bhavin J. Shastri}

\affil[1]{Centre for Nanophotonics, Department of Physics, Engineering Physics, and Astronomy, Queen’s University, Kingston, Ontario, Canada}

\affil[2]{Department of Electrical and Computer Engineering, University of Pittsburgh, Pittsburgh, Pennsylvania 15261, USA }

\affil[3]{Department of Materials Science \& Engineering, University of Maryland, College Park, Maryland 20742, USA }

\affil[4]{Institute for Research in Electronics and Applied Physics, University of Maryland, College Park, Maryland, 20742, USA}

\begin{document}

\title{Cryogenic thermo-optic response of low-loss phase change material for non-volatile photonic phase shifter}

\maketitle

\begin{abstract}
Phase change materials (PCMs) can enable non-volatile optical memory through the large refractive index and extinction coefficient contrast between crystalline and amorphous states. Among them, $\mathrm{Sb}_{2}\mathrm{Se}_{3}$ combines low optical attenuation at telecom wavelengths, making it promising for low-loss programmable phase shifters. Its non-volatility is particularly useful for cryogenic systems, where power dissipation and thermal load constrain scalability. However, its optical properties at cryogenic temperatures remains poorly understood. Here, we report the first cryogenic optical characterization of $\mathrm{Sb}_{2}\mathrm{Se}_{3}$ integrated on a foundry silicon photonic platform from 4K to 300K for both crystalline and amorphous states. We observe that the magnitude of the thermo-optic coefficient decreases upon cooling, whereas the optical attenuation changes only weakly for both the amorphous and crystalline phases. We report the stability of the material upon repeated thermal cycling. These measurements provide the material parameters and stability required to design low-loss, non-volatile photonic memory elements for scalable cryogenic information processing.

\end{abstract}


\section{Introduction}
Qubits have finite coherence times, \cite{simmons2024scalable}. This imposes strict latency constraints on quantum operations at cryogenic temperatures and on the real-time ancillary classical processing including control logic or qubit readout and decoding. Photonic processors are well-suited to these tasks for these time-critical applications because it offers low-latency information processing \cite{shastri2021photonics, wang2025integrated}, but these advantages are diminished when the photonic hardware is located outside the cryostat. Integrating photonics circuits at cryogenic temperatures, inside the cryostat, close to the quantum processor, could reduce communication latency and enable scalable quantum information processing \cite{aharonovich2026programmable, pintus2023cryogenic, eltes2020integrated}. Such integration, however, requires optical components that hold their programmed state without imposing excessive electrical or thermal overhead.

Programming a photonic integrated circuit (PIC) requires setting the states of its memory elements, devices whose refractive index, and hence the optical phase they impart, determine how light propagates through the circuit to implement a desired computation. These states are conventionally set with electrical heaters acting through the thermo-optic effect\cite{kari2023optical, tait2016microring, carolan2015universal}, an approach that is volatile and requires continuous power to maintain. Cryogenic operation compounds the cost in two ways. First, each heater dissipates static power that active cooling must remove to prevent thermal crosstalk, and every electrical feed-through from the cold stage introduces a heat leak that scales with device count whether or not the devices are driven \cite{alam2023cryogenic}. Second, tuning efficiency also decreases because the thermo-optic coefficient of silicon near 4K is more than four orders of magnitude smaller than at room temperature, so a much larger temperature change is required to produce the same phase shift. Alternative volatile tuning mechanisms \cite{alam2023cryogenic, stanfield2019cmos,chakraborty2020cryogenic,lee2020high,gehl2017operation} may reduce device-level dissipation, but each still requires a persistent electrical connection and inherits the same heat-leak scaling problem. A non-volatile photonic memory (a device that holds its state without continuous power) is therefore a prerequisite for large-scale cryogenic photonic systems.

Ferroelectric devices can hold a polarization state without electrical power, but require a constant bias to read from the memory \cite{wen2025ferroelectric}. Mechanical and piezoelectric approaches, including MEMS, NOEMS, and AlN-based devices, can be non-volatile and avoid static dissipation, but may present trade-offs in actuation voltage, footprint, fabrication complexity, or cryogenic cycling reliability \cite{hu2025nonvolatile, wen2024strain}. Here, we show that phase-change materials (PCMs) can enable photonic memory that is programmed at room temperature, read at cryogenic temperatures, and holds its state with no continuous electrical tuning inside the cryostat.

Phase-change materials (PCMs) store information in their structural phase, making them a promising candidate for non-volatile photonic memory. Controlled  energy pulses can reversibly switch PCMs between crystalline and amorphous states, which have distinct refractive indices and absorption \cite{wuttig2017phase, fang2023non, abdollahramezani2020tunable, rahimi2025high}. Amorphization is typically induced by a short (ns to $\mu$s), high-amplitude pulse that melts the material, followed by rapid quenching to suppress crystallization. Conversely, a longer ($\mu$s to ms), lower-amplitude pulse crystallizes the material by maintaining it above its crystallization temperature. Partial amorphization, in which a fraction of the film is switched, yields analog, multilevel states within a single device \cite{abdollahramezani2020tunable,zhang2019miniature}.

Chalcogenide PCMs such as $\mathrm{Ge}_{2}\mathrm{Sb}_{2}\mathrm{Te}_{5}$ (GST), $\mathrm{Ge}_{2}\mathrm{Sb}_{2}\mathrm{Se}_{x}\mathrm{Te}_{5-x}$ (GSST), and $\mathrm{Sb}_{2}\mathrm{Se}_{3}$ exhibit large refractive index contrasts of order unity at telecom wavelengths \cite{sahoo2022gsst,lei2022magnetron,zhang2019broadband}. GST offers a large index contrast, but also shows substantial optical absorption; in quantum photonic circuits, this is not merely an efficiency penalty, since photon loss acts as a decoherence channel and directly caps heralding and success rates. These limitations have motivated interest in low-loss PCMs such as GSST and $\mathrm{Sb}_{2}\mathrm{Se}_{3}$ \cite{prabhathan2023roadmap}. $\mathrm{Sb}_{2}\mathrm{Se}_{3}$ is particularly attractive because its extinction coefficient near 1550nm is $<10^{-5}$ in the amorphous phase rendering it effectively transparent \cite{rios2022ultra, delaney2020new}. Furthermore, $\mathrm{Sb}_{2}\mathrm{Se}_{3}$ can be deposited on foundry-fabricated silicon photonic chips using a back-end-of-line sputtering step  \cite{rahimi2025high}, providing a path for scalable CMOS-compatible platforms \cite{shastri2021photonics}. 

For cryogenic operation, however, non-volatility alone is not sufficient; two further conditions must hold. First, the optical response of the programmed device must shift predictably after cooling. Although the material phase is non-volatile, the refractive indices of both the PCM and the underlying silicon vary with temperature through the thermo-optic effect \cite{frey2006temperature, komma2012thermo}. If these thermo-optic shifts are deterministic and well-characterized, the accumulated optical phase change can be predicted and accounted for at programming time, enabling memories to be written at room temperature and read reliably at cryogenic temperatures. Second, the programmed state must survive thermal cycling. The large thermal expansion mismatch between a chalcogenide film and a silicon substrate admits the possibility of stress-induced structural change, partial recrystallization, or film degradation over repeated cooldowns. Meeting both conditions would eliminate the static power consumption and heat leaks associated with per-device feedthroughs. While the thermo-optic coefficients of some PCMs have been reported near room temperature \cite{zaini2025broadband, stegmaier2016thermo}, the cryogenic optical properties of PCMs is largely unexplored. So far, only GST has any electrical cryogenic data \cite{lombardo2026phase} and optical characterization has been limited to 4K and 300K, excluding the intermediate temperatures \cite{adya2025non}. The significant attenuation contrast in GST limits it to amplitude modulation applications, whereas $\mathrm{Sb}_{2}\mathrm{Se}_{3}$ can provide pure phase shift modulation.

Here, we report the thermo-optic response and thermal-cycling stability of $\mathrm{Sb}_{2}\mathrm{Se}_{3}$, from 4K to 300K — to our knowledge, the first continuous characterization of this low-loss material across this range. $\mathrm{Sb}_{2}\mathrm{Se}_{3}$ was cladded on silicon microring resonators; by isolating the PCM contribution from the well-characterized silicon response, we extract the material refractive-index $n(T)$ and its temperature derivative $\frac{dn}{dT}$.  The $\frac{dn}{dT}$ of the crystalline state changes from $1e-3 K^{-1}$ at 300K to $<1e-5 K^{-1}$ at 4K. The amorphous state experiences a similar change, but with more variation near 300K. The absorption of the material did not change significantly in either phase. We further show that the programmed state is retained over 3 thermal cycles, with the cryogenic effective index reproducible to within 0.0010 for the amorphous state and 0.0023 for the crystalline state. The index contrast between the amorphous and crystalline states was preserved, meaning information can be stored at room temperature and retrieved at cryogenic temperatures, and requires no continuous electrical tuning inside the cryostat.

These results establish the material basis for cryogenic phase-change photonic memory and remove a central obstacle to dense, low-latency, and thermally efficient photonic memory for cryogenic and quantum systems.

\section{Results and Discussion}

\subsection{Cryogenic Transmission Measurement}
\begin{figure}
  \includegraphics[width=\linewidth]{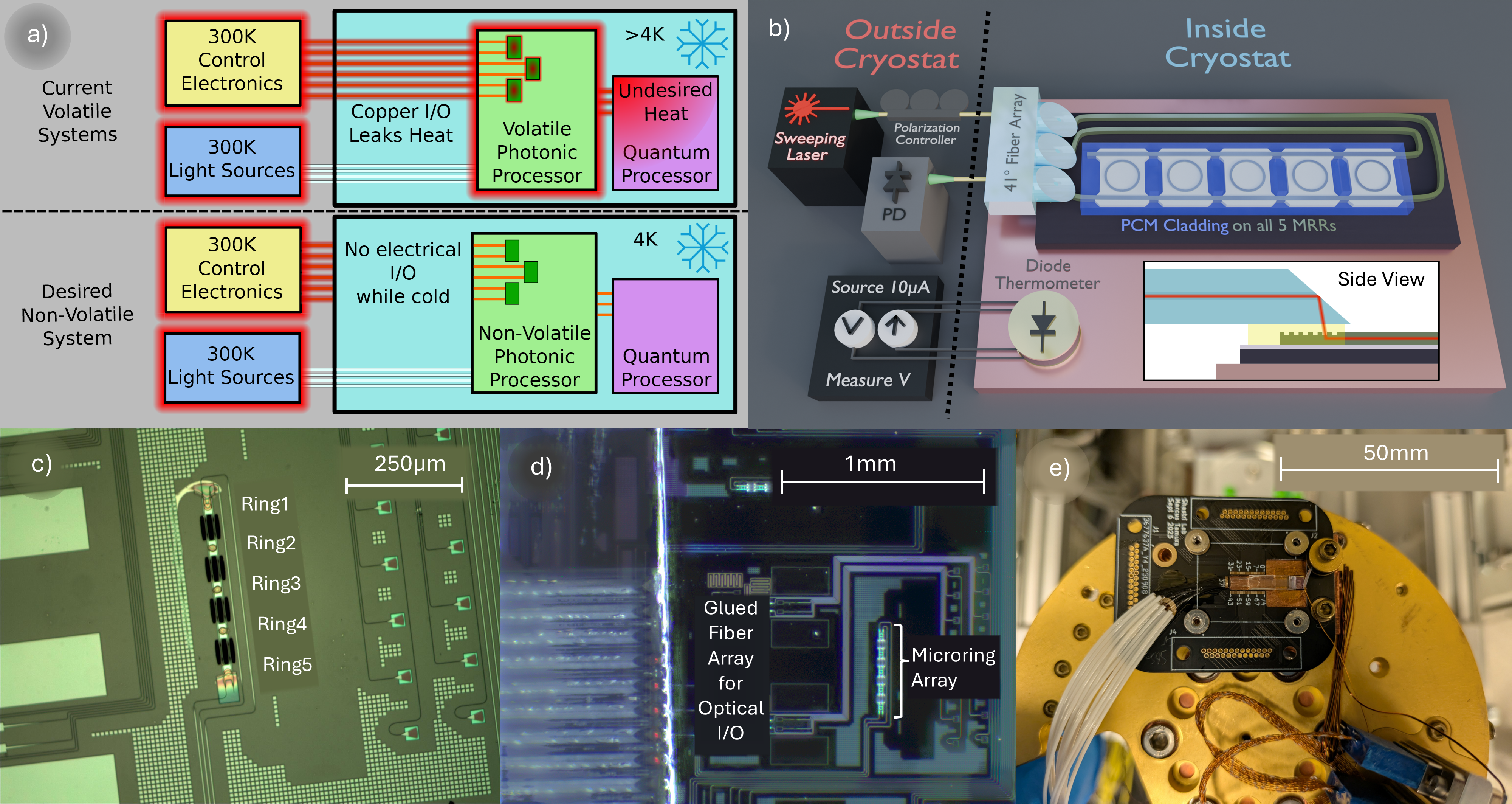}
  \caption{a) Current systems which use volatile methods suffer with higher heat loads due to static power usage and leakage of heat through copper electrical connections from outside the cryostat. Our desire is to use non-volatile materials to remove both sources of heat load for quantum applications. b) A schematic of the experimental setup. A wavelength tunable laser injects light into the chip via a fiber array glued to the PIC above the grating couplers (see side view inset). Light is then filtered by the microrings depending on its wavelength and the temperature. The light is then detected off-chip outside the cryostat with the photodiode (PD). Simultaneously, a diode thermometer measures the temperature as the PIC is cooled down to 4K. c) Optical micrograph of the silicon microrings cladded with 30nm of $\mathrm{Sb}_{2}\mathrm{Se}_{3}$ and 10nm of protective $\mathrm{Si}\mathrm{O}_{2}$. Cladding only covers the rings, with deep trench in between the microrings. Some non-uniformity is observed on Rings 1 and 5. d) Fiber array has glued over top of the grating couplers of a bare silicon chip. e) The package screwed into the 4K stage of cryostat. A circular DT-670 thermometer screwed onto the PCB package to monitor the temperature.}
  \label{Structure}
\end{figure}

The sample was optically packaged (see the Design and Packaging section for more details) then screwed onto the 4K stage of our closed-cycle cryostat. A calibrated DT-670 silicon thermometer diode was tightly screwed directly onto the PCB package to ensure our measurement of the temperature was as close as possible to the photonic integrated circuit (PIC). Optical micrographs of our PIC, its packaging and its placement in the cryostat are shown in Figure \ref{Structure} c) ). 

\begin{figure}
  \includegraphics[width=0.5\linewidth]{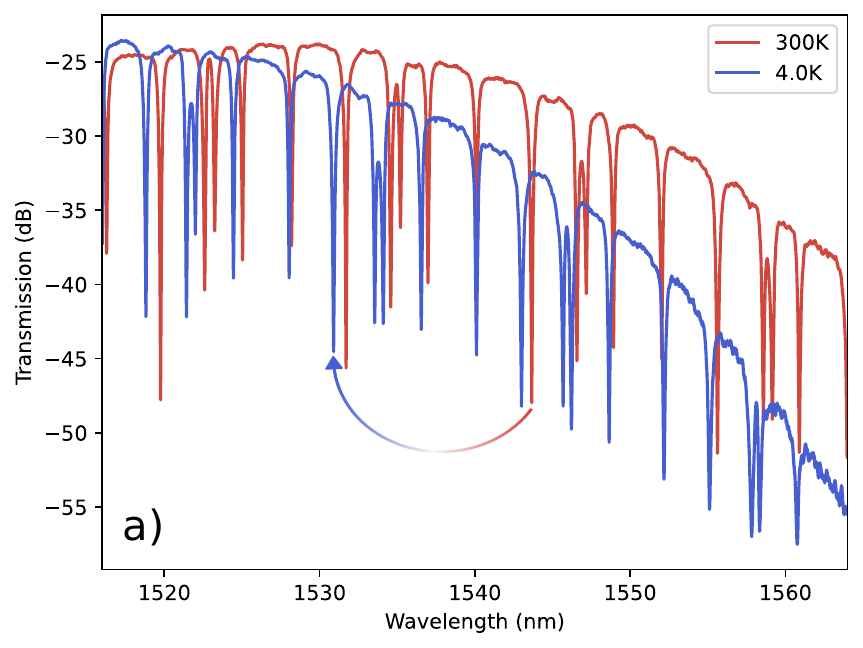}
    \includegraphics[width=0.5\linewidth]{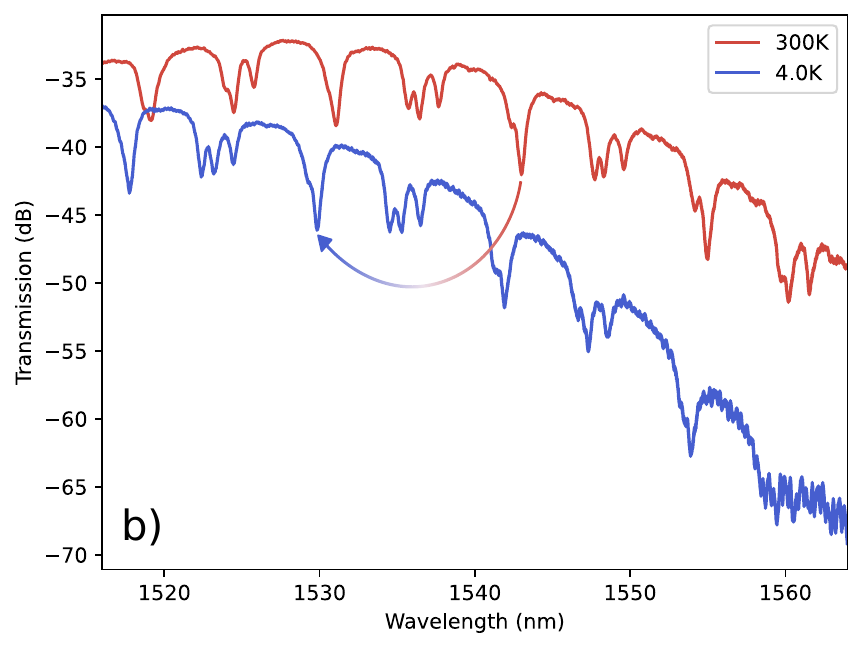}
  \caption{Transmission spectra of the microring weightbank at 300K and 4K while the $\mathrm{Sb}_{2}\mathrm{Se}_{3}$ is a) amorphous and b) crystalline. Two of the resonances of Ring 1 are connected by the curved arrow, indicating a resonant peak shift greater than a full free spectral range. Thermal contraction can slightly change alignment conditions, despite gluing the fiber array directly to the chip, causing an overall change in the grating coupler transmission.  }
  \label{SbSeAmandCrTransRoomCryo}
\end{figure}

A radiation shield was placed around the 4K stage in thermal contact with our 40K stage. Then the system was brought to vacuum. Using our optical spectrum analyzer, the transmission spectrum through the set of microrings was measured from 1516nm to 1564nm. The temperature and optical spectra were captured every 5 seconds for the duration of a cooldown. To prevent any possible ice buildup on the PIC from water that was not successfully removed during vacuum, the 4K stage was heated while also being cooled. This was to allow the 40K stage and radiation shield to cool faster and water would freeze to them before they froze to the 4K stage and the sample package. Once the 40K stage was below the freezing point of water, the heater for the 4K stage was turned off to allow the system to fully cool down. After roughly 2 hours, the system was fully cooled. The cryostat was then turned off and the system was allowed to naturally warm to room temperature over the course of 24 hours. During the warmup experiments, temperature and spectra measurements were taken every 30 seconds. Cooling from 300K to 4K and warming back to 300K formed a full thermal cycle. The transmission spectra shifted by a full free spectral range when cooling from room temperature to 4K, as seen in Figure \ref{SbSeAmandCrTransRoomCryo}.

\begin{figure}[h]
\includegraphics[width=\linewidth]{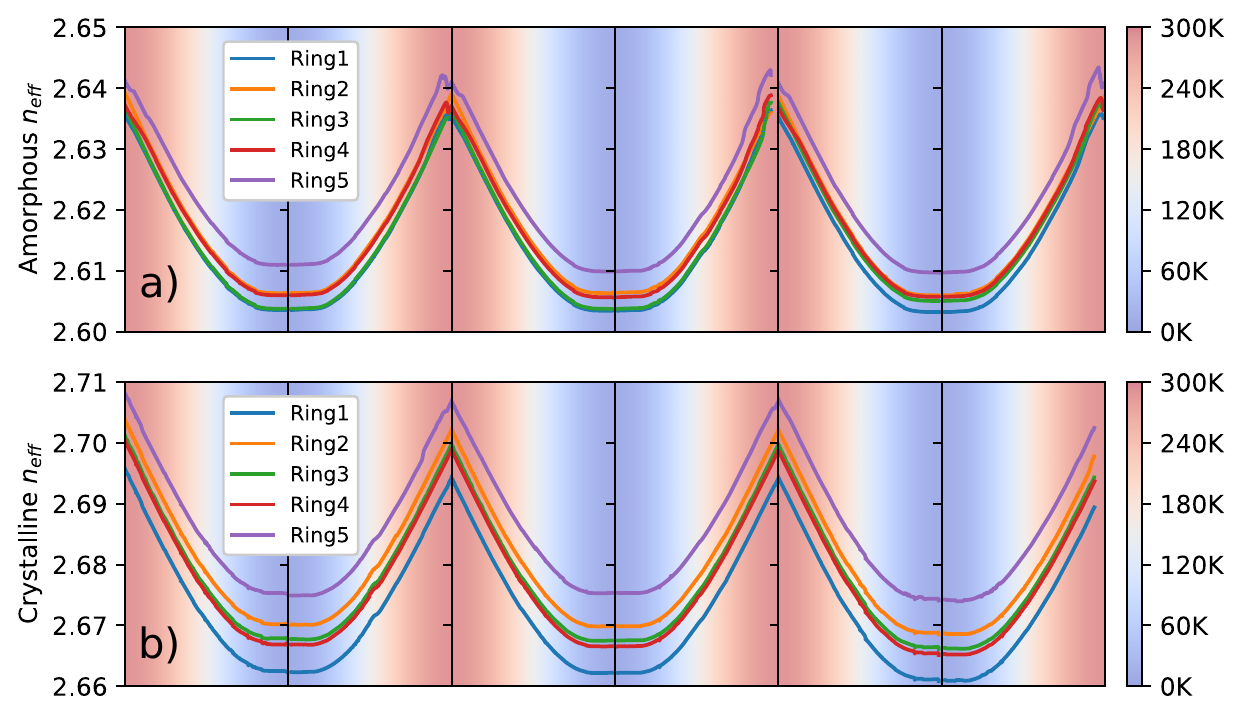}
  \caption{Effective refractive index of the silicon waveguide cladded with a) amorphous and b) crystalline $\mathrm{Sb}_{2}\mathrm{Se}_{3}$ under repeated thermal cycles}
  \label{SbSeAm_and_CrEff}
\end{figure}

Three thermal cycles were performed on the chip with the as-deposited amorphous $\mathrm{Sb}_{2}\mathrm{Se}_{3}$. The chip was then heated to $200\degree\text{C}$ on a hotplate to crystallize the $\mathrm{Sb}_{2}\mathrm{Se}_{3}$. Three more thermal cycles were then performed. The transmission spectra data was fitted to a microring model \cite{bogaerts2012silicon} to retrieve the effective index changes of the microrings as shown in Figure \ref{SbSeAm_and_CrEff}. The details of this analysis are provided in the Methods section.

\begin{table}[]
    \centering
    \begin{tabular}{|c|c|c|c|c|c|c|}
        \hline
          & Ring1 & Ring2 & Ring3 & Ring4 & Ring5 & Simulated\\
         \hline
        $\Delta n_{eff}$ & 0.0560 & 0.0630 & 0.0615 & 0.0575 & 0.0635 & 0.0638\\
        \hline 
    \end{tabular}
    \caption{The change in the effective index at room temperature after annealing, converting the amorphous $\mathrm{Sb}_{2}\mathrm{Se}_{3}$ into crystalline. The simulated shift was determined using FEMWELL \cite{gehring2025femwell}.}
    \label{tab:Am2Cr}
\end{table}

\subsection{Refractive Index Analysis}

The effective refractive indices decreased with temperature, but tended to reach a plateau below 60K (see Figure \ref{SbSeAm_and_CrEff}). The phase state of the phase change material was well preserved for all rings across all cooldowns. Generally the cooldown and warmup was repeatable but often slight differences emerged, especially during the warm up near 250K to 300K. We believe this is partially due to some relaxation of stress incurred from differing thermal expansion coefficients between the materials and the large temperature changes. Ring 5 was particularly prone to this, and from the microscope image in Figure \ref{Structure} a) we suspect its coverage with PCM was less well-behaved compared to the other rings. This varied near-room temperature behavior did not appear  as significantly while in the crystalline state.

\begin{figure}
  \includegraphics[width=\linewidth]{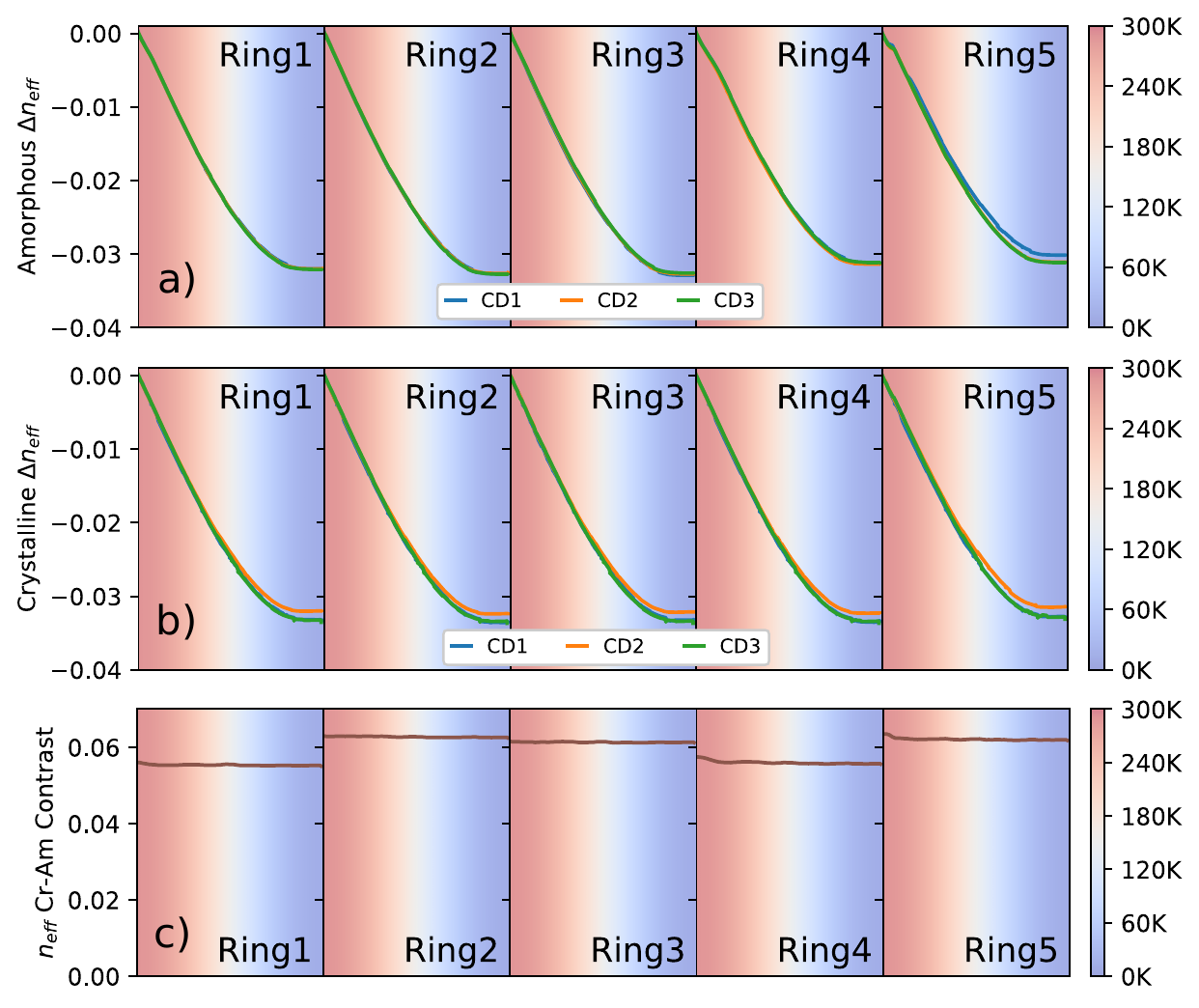}
  \caption{Change in the effective refractive index from its value at 300K for the silicon waveguides cladded with a) amorphous and b) crystalline $\mathrm{Sb}_{2}\mathrm{Se}_{3}$ under repeated thermal cycles c) The effective refractive index contrast between crystalline and amorphous states. Data presented is the mean across all cooldowns. Despite some variation between rings, contrast is maintained across the whole temperature range.}
  \label{SbSeAmandCrDeltaEff}
\end{figure}

\begin{figure}
  \includegraphics[width=\linewidth]{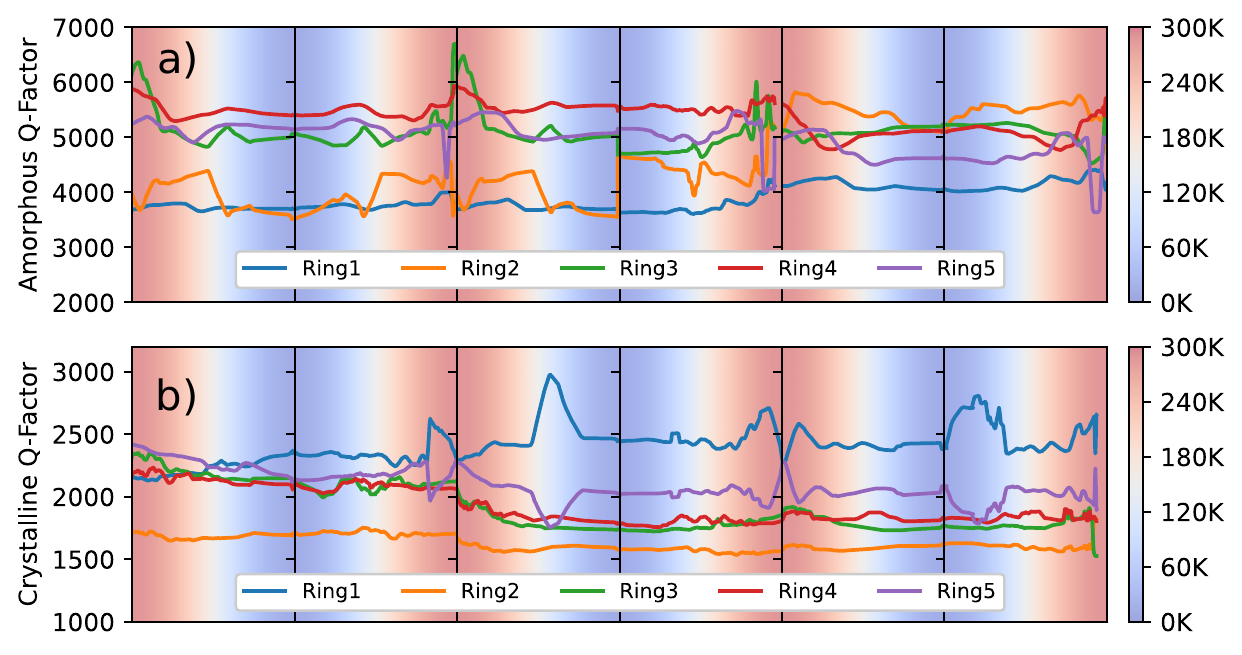}
  \caption{Quality for the microrings cladded with a) amorphous and b) crystalline $\mathrm{Sb}_{2}\mathrm{Se}_{3}$ under repeated thermal cycles}
  \label{SbSeQfactor}
\end{figure}

As is typical with microrings \cite{mirza2024experimental}, the average effective indices differed slightly between different rings due to fabrication variation. For this reason, we also plot the change in effective refractive index in Figure \ref{SbSeAmandCrDeltaEff} taking 300K as our origin. Here an interesting trend emerges, where the change in effective refractive index is extremely consistent between cooldowns. Comparing between rings, there is some small deviation still for the amorphous state. For the crystalline state, the rings were remarkably consistent between themselves, excluding ring 5. We observed some changes in the quality factor and the extinction ratio between room temperature and cryogenic conditions (see Figure \ref{SbSeQfactor} and 
Table \ref{tab:extinction}). However, no clear trend emerged. When ring resonances were close together and overlapping it is more difficult to determine the width of each individually. This was especially true for Ring 1 and Ring 5 for the crystalline data, which from the transmission spectra it is difficult to tell that they are in fact two separate resonances. This is the reason there were some fitting artifacts in the quality factor plots where one ring will inversely correlate with another. The limitation on the quality factors are due high over-coupling to the rings and is not due to the loss of the $\mathrm{Sb}_{2}\mathrm{Se}_{3}$ material.

\begin{table}[]
    \centering
    \begin{tabular}{|c|c|c|c|c|c|}
        \hline
        Extinction Ratio  & Ring1 & Ring2 & Ring3 & Ring4 & Ring5\\
         \hline
        Before PCM deposition 300K & 24.99dB & 22.73dB & 26.85dB & 25.18dB & 25.66dB \\
        \hline 
        Amorphous $\mathrm{Sb}_{2}\mathrm{Se}_{3}$ 300K & 21.33dB & 17.18dB & 10.87dB & 14.80dB & 13.00dB \\
        \hline
        Amorphous $\mathrm{Sb}_{2}\mathrm{Se}_{3}$ 4K & 18.44dB & 15.22dB & 14.63dB & 12.18dB & 14.12dB \\
        \hline
        Crystalline $\mathrm{Sb}_{2}\mathrm{Se}_{3}$ 300K & 5.94dB & 4.19dB & 4.79dB & 3.07dB & 4.17dB \\
        \hline 
        Crystalline $\mathrm{Sb}_{2}\mathrm{Se}_{3}$ 4K & 5.97dB & 4.62dB & 3.92dB & 2.95dB & 3.92dB \\
        \hline 
    \end{tabular}
    \caption{The extinction ratio of each of the rings at various temperatures and phase states.}
    \label{tab:extinction}
\end{table}

\begin{figure}
  \includegraphics[width=0.9\linewidth]{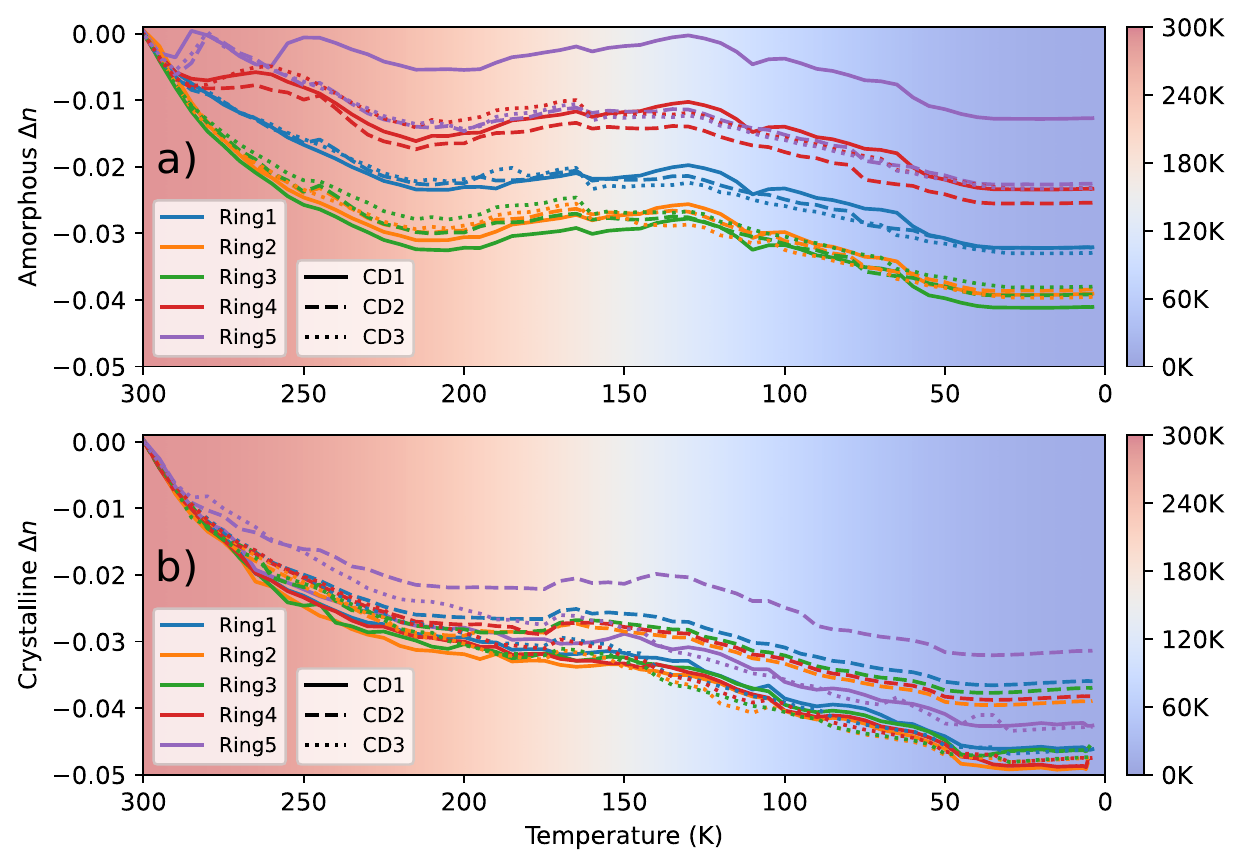}
  \caption{Change in the material refractive index from its value at 300K for a)amorphous and b) crystalline $\mathrm{Sb}_{2}\mathrm{Se}_{3}$ under repeated thermal cycles}
  \label{SbSeAmandCrDeltaMat}

  \includegraphics[width=0.9\linewidth]{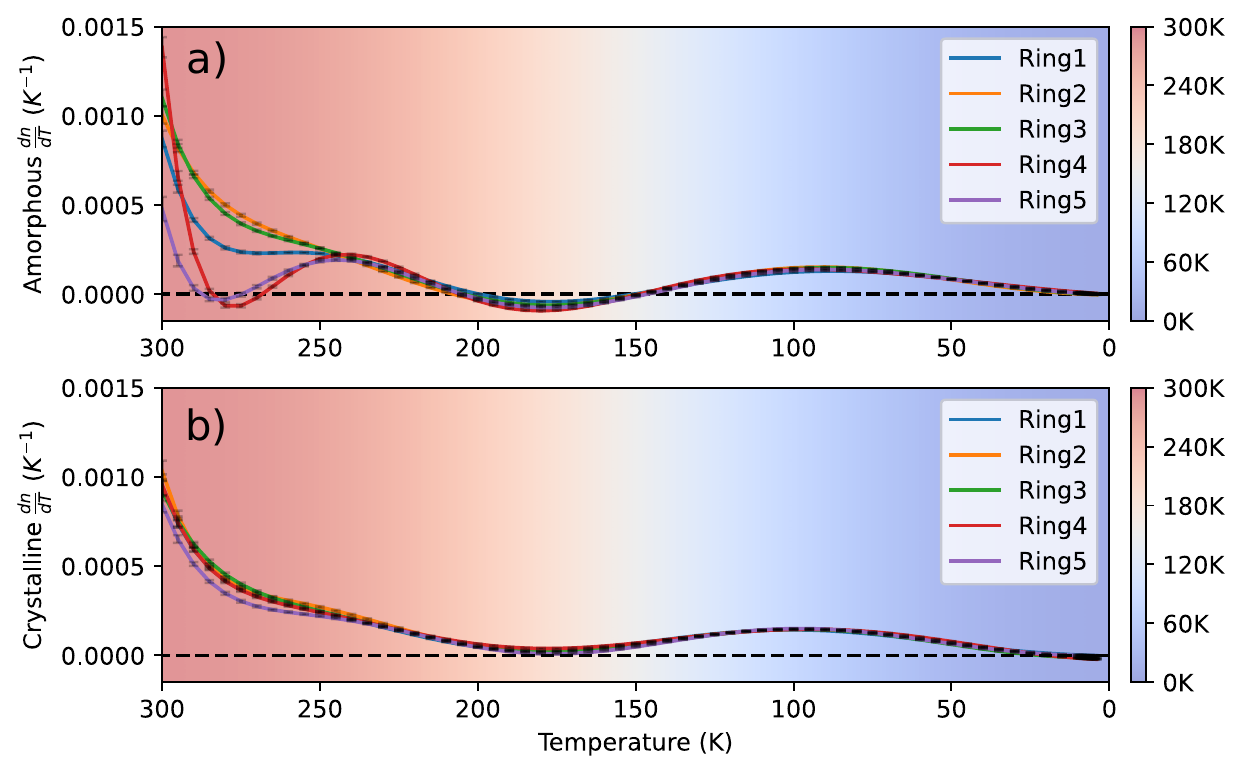}
  \caption{Thermo-optic coefficient for a) amorphous and b)  $\mathrm{Sb}_{2}\mathrm{Se}_{3}$. The mean of each of the three cooldowns was taken and a 11th order polynomial fit was made from the mean to extract the general trend. The derivative of that polynomial fit is plotted. The error bars represent the error from performing the polynomial fit to smooth the data.}
  \label{SbSeAmandCrThermoOptic}
\end{figure}

These effective refractive index measurements are only useful for this system, since they dependent on geometry. In order to generalize our data for applications beyond this specific geometry, the material refractive index must be extracted by accounting for the proportions of the refractive index shifts due to the silicon versus the $\mathrm{Sb}_{2}\mathrm{Se}_{3}$. This analysis is detailed in the Methods section, but the results are shown here in Figures \ref{SbSeAmandCrDeltaMat}.

The material refractive index also decreased as the temperature decreased. Significant variation was observed between the rings for the amorphous data. There are a few possible explanations for this. Since the amorphous state is a random disordered state, it stands to reason that each ring may have a different random disordered state as its cladding. While all of them are still amorphous, there may be some variation between what random disordered state exists over each ring. Another possible explanation is if each ring has a slightly different waveguide geometry (slightly wider silicon waveguides, thicker cladding, etc. ). Cladding the entire ring with PCM was an intentional experimental design choice to help ensure a uniform thickness of PCM. Still, some variation is expected \cite{mirza2024experimental}, but based on the crystalline data, it seems more likely that the difference between rings may be due to the former cause. The crystalline material refractive index values were much less varied between microrings. If different waveguide geometries between the rings was the cause, we would expect to see the rings behave similarly between their amorphous and crystalline states. Instead, each crystallized ring behaves similar to each other, with the exception of perhaps ring 5. All of the material refractive index data showed a small maxima near 150K. Further investigation may be warranted near this temperature range.

We observed that the thermo-optic coefficient of $\mathrm{Sb}_{2}\mathrm{Se}_{3}$ does decrease significantly at cryogenic temperatures for both the crystalline and amorphous state. In the 250K to 300K, we observed variability in the thermo-optic effect between the rings. Below this range, the rings tended to act similarly in response to temperature changes. We are not certain of the cause of this, but hypothesize that it could be due to mechanical shifting of the $\mathrm{Sb}_{2}\mathrm{Se}_{3}$ and silicon during cooldown. Silicon and $\mathrm{Sb}_{2}\mathrm{Se}_{3}$ have different thermal expansion coefficients ($2.32\times 10^{-6}K^{-1}$ \cite{ekin2006experimental} and $\approx 1.2\times 10^{-5}K^{-1}$ depending on the direction \cite{herrmann2020lattice}) which may be causing stress induced refractive index changes. This variability was only seen for the amorphous state, as the crystalline data was overall more consistent between rings. We suspect this could be due to slightly different random states in the disordered structure between the rings. 

Materials can have different properties comparing thin-film and bulk. A primary goal of this paper is to demonstrate how the thin-film phase change material behaves at cryogenic temperatures on a silicon photonics foundry platform. This allows us to better understand how the material would behave for more typical applications such optical memory which would want the scalability of silicon. However, this experimental method could be improved upon in future work. Since we measure the resonance peak positions, it is also difficult to extract any change in the dispersion of the material between the resonance peaks. Currently, since we are using microrings, any birefringent behavior of the material is challenging to observe. Additionally, silicon's own thermo-optic changes make up the majority of our observed effective index changes during cooling. Other photonic materials such as silicon nitride with an order of magnitude smaller thermo-optic coefficient compared to silicon \cite{stegmaier2016thermo, elshaari2016thermo}, could allow for more precision in this type of measurement. Nevertheless, our experimental conditions more closely mimic the typical usage case of cryogenic optical memory than past bulk sample cryogenic measurements which can achieve better precision overall \cite{komma2012thermo}.

Microrings are remarkably sensitive devices, requiring very little refractive index change to induce large resonance peak shifts and therefore large transmission changes. This means optical memory applications may not need to fully switch between amorphous and crystalline states to be capable of tuning from full to null transmission. PCMs have been shown to be capable of holding multibit memory by switching portions of the material rather than switching the entire material. This is accomplished through precise control of memory setting thermal pulses. Due to the higher reliability of our crystalline data, we suspect that future cryogenic memories primarily stay in the mostly crystalline state space when setting multibit memories. The change in the effective indices appears more deterministic. The small deviations in the effective indices between cooldowns could limit the bit resolution of the cryogenic memory and could require additional low temperature switching approaches to fine tune \cite{adya2025non}.

\section{Conclusion}
We have experimentally shown that $\mathrm{Sb}_{2}\mathrm{Se}_{3}$ has a decreased thermo-optic coefficient at cryogenic temperatures in both the amorphous and crystalline states at telecom wavelengths on a silicon photonic foundry platform. $\mathrm{Sb}_{2}\mathrm{Se}_{3}$ is used for non-volatile optical phase shift modulation. The amorphous state exhibited more variability in the thermo-optic coefficient comparing between rings than the crystalline state near room temperature, but this variability disappeared at cryogenic temperatures. We believe this may be due to small morphological changes from the stress induced from thermal contraction. We also observed no significant change in the quality factor comparing cold to warm, suggesting the attenuation coefficient change is small. The phase of matter of the material did not appear to change at all even under repeated thermal cycles from 300K to 4K and back again. Extracting the material refractive index changes from the geometry dependent effective refractive indices was accomplished through the use of FEM simulations. We hope this material data will help the development of cryogenic optical memories for quantum information architectures. 
\externaldocument{supplementary}
\section{Methods}
\subsection{Design and Packaging}
The photonic integrated chips (PICs) were fabricated by Advanced Micro Foundry (AMF). Five microrings of radii 8.0000$\mu$m, 8.01213$\mu$m, 8.02426$\mu$m, 8.03639$\mu$m 8.04852$\mu$m were arranged in series on two separate PICs. The first PIC used microrings composed of bare uncladded silicon rib waveguides. The other one used identical silicon rib waveguides, but had sputtered $\approx$29.12nm of $\mathrm{Sb}_{2}\mathrm{Se}_{3}$, and 10nm of silicon dioxide to protect the PCM from incidental oxidation. Then the chips were ready for optical packaging.

Printed circuit boards with cutouts were screwed onto copper mounting plates. The chips were glued to copper plates (Elmer's rubber cement) within the cutouts. The copper plate was loaded into our temperature controlled optical probe state. A glass v-groove fiber array was aligned to grating couplers on the PIC. The fiber arrays were polished at a 41 degree angle so input light would be sent by total internal reflection down towards the PIC \cite{mckenna2019cryogenic}. The grating couplers were optimized to accept C-band light at 8 degrees from the vertical. After optical alignment at room temperature, optical alignment was repeated at 80\degree C. The fiber array was then lifted in the vertical direction and two-part thermal epoxy (Epoxy Technology 301-2) was spread on the bottom of the fiber array. The fiber array was brought down, and the thermal epoxy was squished between the fiber array and the chip. The thermal epoxy was allowed to cure over 3 hours. Then more thermal epoxy was applied between the PCB and the fibers coming from the back end of the v-groove array. This second set of gluing was again let to cure over 3 hours. The clamp holding the v-groove in place was then released and the package was allowed to cool back to room temperature.

\subsection{Analysis and Fit}
The measured transmission spectra of the weightbank are combined with the transmission spectra of the grating couplers. To isolate the transmission spectra of just the weightbank, a low pass filter is applied to the measured transmission spectra to smooth the small Fabry-Perot noise. A find peaks algorithm is used, and points in between the peaks should be pure background. A linear interpolation is made between each point in between the peaks to approximate the grating coupler background. This background spectra is subtracted from the original raw measured transmission spectra to give a spectra of just the microring weightbank and some Fabry-Perot noise. This is the spectra that we fit our microring weightbank model against.

The transmission  of a symmetrically coupled add-drop microring is given by 
\begin{equation}
T_{thru}(\lambda)=\frac{r^2(a^2-2acos(\beta(\lambda) 2 \pi R) + 1)}{1-2r^2acos(\beta(\lambda) 2 \pi R)+r^4a^2},
\label{transmissioneqn}
\end{equation}

where $r$ is the amplitude self-coupling coefficient, $a$ is the  single-pass amplitude transmission, and $R$ is the microring radius \cite{bogaerts2012silicon}. The propagation constant, $\beta$ is a function of wavelength ($\lambda$) and the effective refractive index ($n_{eff}$) of the microring fundamental mode given as

\begin{equation}
\beta(\lambda)=\frac{2\pi n_{eff}(\lambda)}{\lambda}.
\label{propagation constant}
\end{equation}

To determine the effective refractive indices dependence on wavelength we use a finite element method eigenmode solver called FEMWELL \cite{gehring2025femwell}. The material refractive indices used in the solver were taken from literature where cryogenic data existed \cite{komma2012thermo, frey2006temperature, leviton2006temperature} and room temperature data was used where no cryogenic data exists \cite{nobile2023nonvolatile}. We found that the effective index dependence is very close to linear with respect to wavelength (see Figure \ref{dispersionwaveguides} in the Supplementary Information), and thus we approximate as

\begin{equation}
n_{eff}(\lambda)\approx n_{eff@1550nm}+\frac{dn_{eff}}{d\lambda}(\lambda-1550nm).
\label{effective_n_wavelength}
\end{equation}

We take the measured transmission spectra with the background removed and fit it to 

\begin{equation}
T_{weightbank}(\lambda) = \prod_{i=1}^{5}T_{thru_i}(\lambda),
\label{total_transmission}
\end{equation}

with the effective refractive indices at 1550nm and self-coupling coefficients of each ring as fit parameters. The single-pass amplitude transmission is kept as a constant hyperparameter. The slope of Equation \ref{effective_n_wavelength} is determined through the FEMWELL simulations. 

In order to extract the material refractive index change as the temperature changes, we must account for the thermo-optic effect of the silicon as well. The change in the effective refractive index is affected by the change in the material coefficients of the silicon and the $\mathrm{Sb}_{2}\mathrm{Se}_{3}$, as given by

\begin{equation}
\Delta n_{eff}(T) = \frac{\partial n_{eff}}{\partial n_{Si}}\Delta n_{Si}(T) + \frac{\partial n_{eff}}{\partial n_{\mathrm{Sb}_{2}\mathrm{Se}_{3}}}\Delta n_{\mathrm{Sb}_{2}\mathrm{Se}_{3}}(T).
\label{effective_n_partial}
\end{equation}

Using FEMWELL, we can extract the partial derivative factors by determining the confinement factors of the light within silicon and the $\mathrm{Sb}_{2}\mathrm{Se}_{3}$. The thermo-optic effect of silicon dioxide is relatively small, even accounting for the large temperature change \cite{leviton2006temperature}. Additionally most of the light is confined within the silicon and the $\mathrm{Sb}_{2}\mathrm{Se}_{3}$ so the contribution of the silicon dioxide can safely be ignored. The thermo-optic change in silicon can be retrieved by cooling an uncladded silicon microring chip. We rearrange Equation \ref{effective_n_partial} to isolate the material refractive index change of $\mathrm{Sb}_{2}\mathrm{Se}_{3}$, given by

\begin{equation}
\Delta n_{\mathrm{Sb}_{2}\mathrm{Se}_{3}}(T)=\frac{\Delta n_{eff}(T) - \frac{\partial n_{eff}}{\partial n_{Si}}\Delta n_{Si}(T)}{\frac{\partial n_{eff}}{\partial n_{\mathrm{Sb}_{2}\mathrm{Se}_{3}}}}.
\label{effective_n_partial_isolated}
\end{equation}

\medskip
\textbf{Supporting Information} \par 
Supporting Information is available from the Wiley Online Library or from the author.

\medskip
\textbf{Acknowledgements} \par 
We acknowledge the support of the Natural Sciences and Engineering Research Council of Canada (NSERC) through the CGS-D program. We would like to acknowledge CMC Microsystems, manager of the FABrIC project funded by the Government of Canada, for the provision of products and services that facilitated this research.

\medskip

%

\bibliographystyle{unsrt}
\bibliography{bib}

\pagebreak
\begin{center}
\textbf{\huge Supplemental Materials: Cryogenic thermo-optic response of low-loss phase change material for non-volatile photonic phase shifter}
\end{center}

\setcounter{figure}{0}
\renewcommand{\figurename}{Fig.}
\renewcommand{\thefigure}{S\arabic{figure}}

\section{Transmission analysis additional details}
\begin{figure}[h]
  \includegraphics[width=0.9\linewidth]{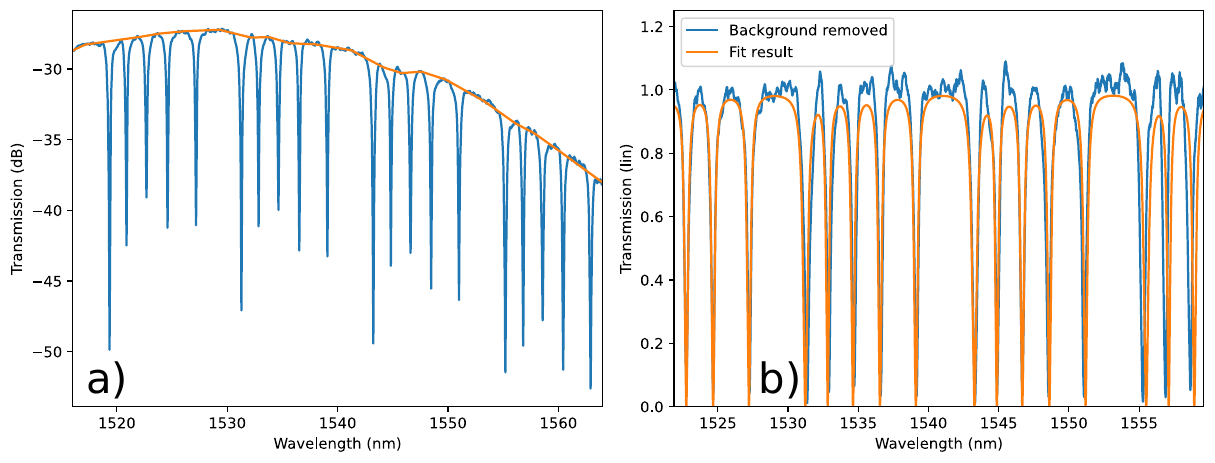}
  \caption{a) A representative example of determining the background grating couplers transmission. b) A representative example of a fit result after having removed the grating coupler (GC) background transmission.  }
  \label{fig:background}
\end{figure}

In order to fit to Equation \ref{total_transmission} in the main text, we must first remove the grating coupler transmission profile. This is done by first taking low-pass filter to the spectra to decrease the impacts of small Fabry-Perot effects. Then we find points in between the resonance peaks, and linearly interpolate between these peaks. The result of this gives an approximation of the grating coupler transmission, as seen in Figure \ref{fig:background}.

Once the grating coupler transmission is identified, it can be subtracted from the raw unfiltered data and converted to a linear scale. This is plotted in Figure \ref{fig:background} b). We fit this spectrum to Equation \ref{total_transmission}, with the fit also being shown in Figure \ref{fig:background} b).

\begin{figure}[h]
  \includegraphics[width=0.9\linewidth]{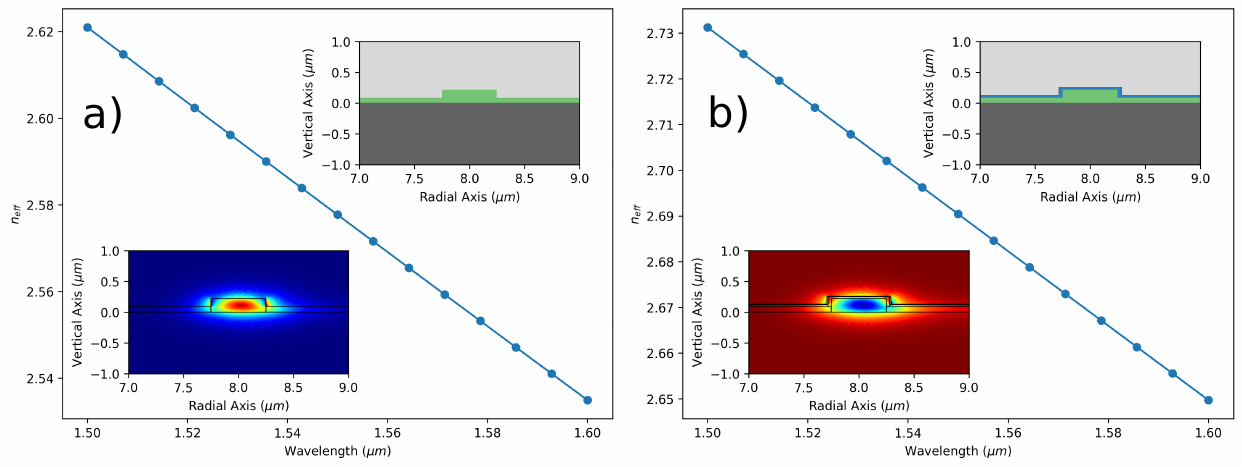}
  \caption{Effective refractive index calculated from a FEMWELL simulation of a 500nm x 220nm silicon rib waveguide with a 90nm thick side slab. a) Air cladded and b) cladded with 30nm of $\mathrm{Sb}_{2}\mathrm{Se}_{3}$, and 10nm of  $\mathrm{Si}\mathrm{O}_{2}$. Both are on a 3um thick buried $\mathrm{Si}\mathrm{O}_{2}$ layer, on a Si handle layer. The effective index changes with wavelength effectively linearly.}
  \label{dispersionwaveguides}
\end{figure}

The wavelength dependence of the effective refractive index is crucial for matching the experimental and fitted free spectral range (FSR). The wavelength dependence was retrieved from an eigenmode solver (FEMWELL) simulations to retrieve the mode effective refractive indices as a function of wavelength. This was performed for a pure silicon rib waveguide, a silicon rib waveguide with amorphous $\mathrm{Sb}_{2}\mathrm{Se}_{3}$ cladding, and another with crystalline cladding. The simulations with PCM also included the protective oxide over the PCM as well to match experimental conditions. Additionally, a radius of 8$\mu$m was chosen to match the microring radii. From these simulations we get a linear dependence of the effective refractive index with respect to the wavelength see Figures \ref{dispersionwaveguides} a) and b). When performing the fits, we used the slope from the linear dependence, but kept the effective refractive index as a fit parameter for each ring. This is to compensate for random fabrication variation, but we assume that the fabrication variation has a weak enough effect on the wavelength dependence that it can be ignored. From how close our fitted results of the effective refractive index and FSR match simulation to experiment we find this approximation appropriate.

\newpage
\newpage

\newpage

\section{Material refractive index extraction additional details}

In order to extract the material refractive index of the $\mathrm{Sb}_{2}\mathrm{Se}_{3}$, we needed to remove the thermo-optic contribution of the silicon part of the waveguide. By cooling down a set of microrings with no PCM cladding, we can extract the silicon material refractive index. We see that it closely but not exactly follows results seen in literature for bulk silicon. The exact cause of the discrepancy is not known, but we suspect slight differences in the materials are the cause. The literature examined a bulk sample, which may also be the reason. Some variation was observed between the rings for these experiments as well. Regardless, our measured silicon results were what was used to extract the $\mathrm{Sb}_{2}\mathrm{Se}_{3}$ material refractive index data since our pure silicon data should be closer in an experimental match.

\begin{figure}[h]
  \includegraphics[width=0.8\linewidth]{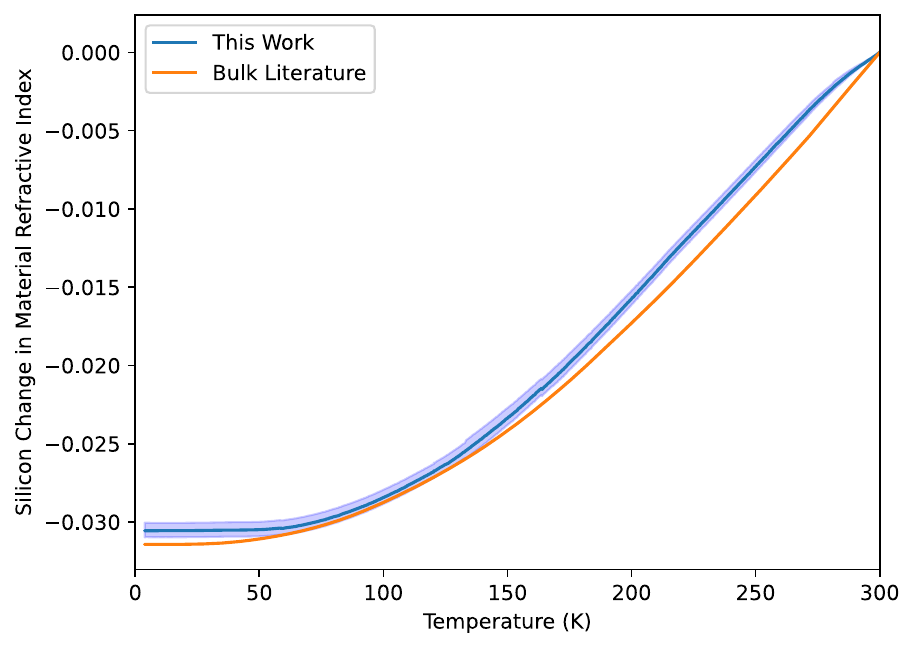}
  \caption{Measured change in the material refractive index of our bare uncladded silicon waveguides compared with against bulk sample reported in the literature \cite{komma2012thermo}. The solid blue curve represents the mean of the change, with the semi-transparent band representing the range between the rings.}
  \label{Si_pure_comparison}
\end{figure}

\newpage
\section{Fitting routine uncertainties}

\begin{figure}[h]
  \includegraphics[width=0.7\linewidth]{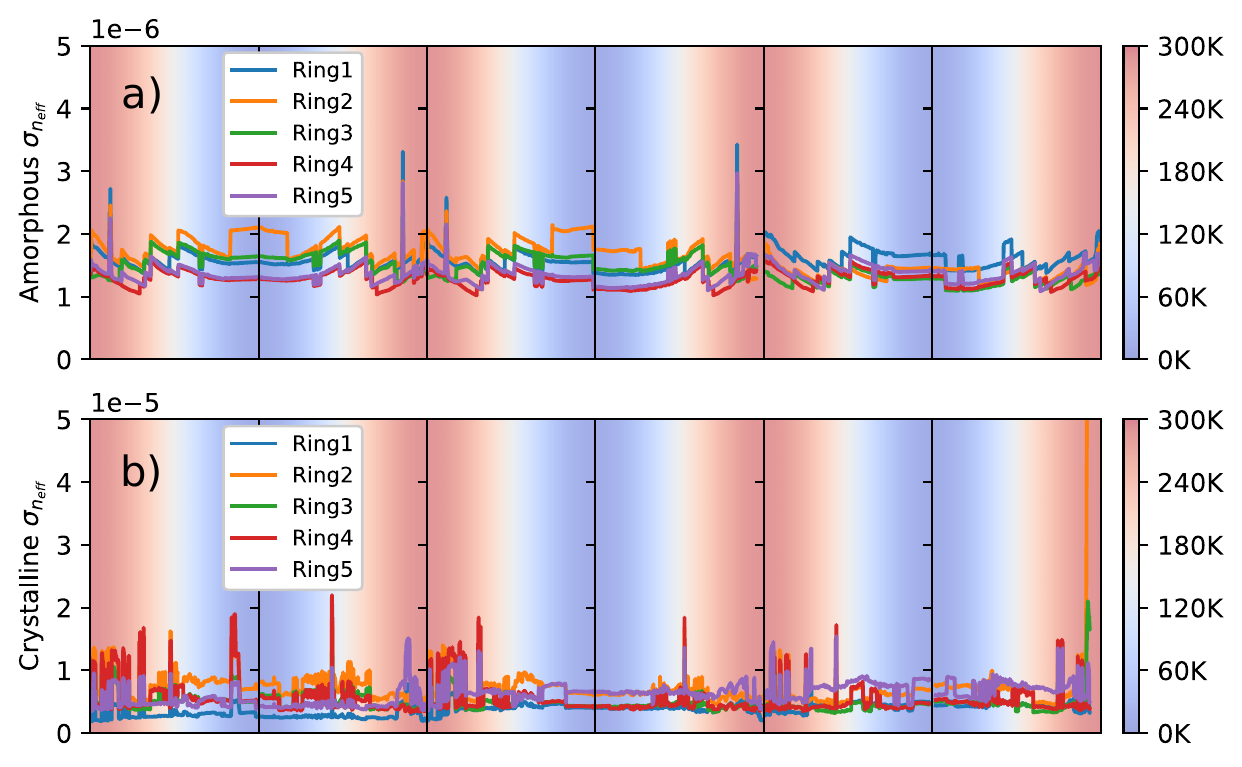}
  \caption{Standard deviation of the effective refractive index retrieved from the fit of the a) amorphous and b) crystalline data.}
  \label{neffStandardDeviation}

  \includegraphics[width=0.7\linewidth]{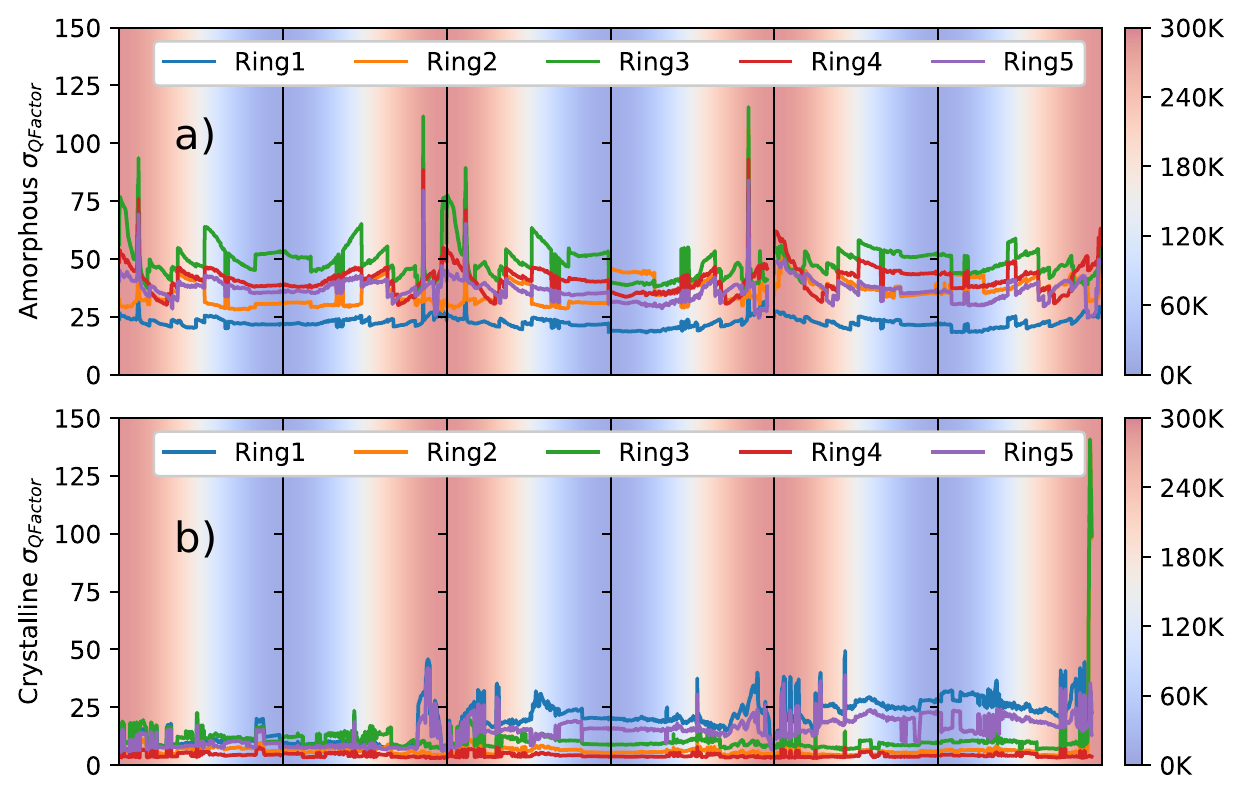}
  \caption{Standard deviation of the quality factor retrieved from the fit of the a) amorphous and b) crystalline data.}
  \label{QfactorStandardDeviation}
\end{figure}

The fitting routine does provide some uncertainty in the fit parameters it finds. These are plotted here in Figures \ref{neffStandardDeviation} and \ref{QfactorStandardDeviation}. Overall, we found that the fitting uncertainty of the effective refractive index to be small, likely because the peaks are so sharp.

\end{document}